\documentclass{knawproc}
\usepackage{psfig}

\def\wisk#1{\ifmmode{#1}\else{$#1$}\fi}

\def\civ{C~{\sc iv}}

\def\lya{Ly\,$\alpha$}

\def\zemzabs{$z_{\rm abs} \approx z_{\rm em}$}

\def\kms{km\,s$^{-1}$}

\begin{document}

% The ``opening'' environment takes care of title, author and headlines
\begin{opening}

\title{Lyman-$\alpha$ absorption: links between CSS quasars and
high-redshift radio galaxies}

\author{Joanne C. Baker}
\addresses{%
MRAO, Cavendish Laboratory, Madingley Rd, Cambridge CB3 0HE, UK\\
}
\runningtitle{Links between CSS quasars and HZRGs}
\runningauthor{J.C. Baker}

\end{opening}

%%%%%%%%%%%%%%%%%%%%%%%%%%%%%%%%%%%%%%%%%%%%%%%%%%%%%%%%%%%%%%%%%%%%%%%%%%%%%%%

\begin{abstract}

Spectroscopy of high-redshift compact steep-spectrum (CSS) quasars 
shows a high incidence of Ly-$\alpha$ absorption close to the quasar redshift 
(\zemzabs). Associated absorption systems like these are rare in 
other types of radio sources, with the notable exception of high-redshift 
radio galaxies (HZRGs). CSS quasars and HZRGs share many properties; 
the hypothesis that they are intrinsically similar objects is presented
and discussed.

\end{abstract}

%%%%%%%%%%%%%%%%%%%%%%%%%%%%%%%%%%%%%%%%%%%%%%%%%%%%%%%%%%%%%%%%%%%%%%%%%%%%%%%

\section{Introduction}

A substantial population of young radio sources is expected, given the
expansion of radio sources and the short-lived
nature of radio emission (Alexander \& Leahy 1987).
This r\^ole may be filled by a class of intrinsically 
small objects, compact, steep-spectrum 
(CSS) sources (as a guide, $l<30$ kpc and $\alpha >0.5$ for $S_{\nu} 
\propto \nu ^{-\alpha}$). Evidence that some CSS sources are indeed 
young is mounting (Readhead et al. 1996), 
although the class may also include frustrated sources whose jets 
are constricted by an unusually dense and/or clumpy interstellar 
medium (ISM).  

The small size of CSS sources means that radio observations probe the
dense nuclear environment of the host galaxies. Absorption by H\,{\sc i} 
gas, perhaps from a dense parsec-scale disk, has been reported towards 
the compact radio components of several CSS sources (Conway 1996). 
Radio depolarisation is also common (Garrington et al. 1991). 
Evidence for jet-ISM interactions  in CSS sources is strong, including 
bent, knotty jets and aligned optical emission (de Vries et al. 1997). 

Evidence that CSS sources differ optically from larger
sources, however, is only now emerging. New differences 
have been found for CSS quasars, including very red optical 
continua, large Balmer decrements and large narrow-line 
equivalent widths which are suggestive of reddening (Baker \& Hunstead 1995). 
One particularly interesting characteristic is that associated 
absorption by enriched H\,{\sc i} gas is exceptionally common 
in CSS quasars (Baker \& Hunstead 1996). In a recent study of the 
complete Molonglo Quasar Sample
(MQS) absorption with \zemzabs\ was found in 80\% of CSS quasars 
with $z>1.4$, where \lya\ and/or \civ\ was visible.

In this short contribution, I will show that CSS quasars share many
similarities with high-redshift radio galaxies (HZRGs), and are arguably
instrinsically similar objects.  This is based largely on the results 
of intermediate resolution spectroscopy (with R. Hunstead) of \zemzabs\ 
absorbers towards high-redshift CSS quasars from the MQS.

\section{Spectroscopy of CSS quasars}

Spectroscopy at 1-\AA\ resolution has been carried out for seven high-$z$ 
CSS quasars drawn from the 408-MHz selected MQS (Baker 1997) with
the Anglo-Australian Telescope. The Ly-$\alpha$ emission-line region 
has been observed for three CSS quasars with $z>2$. 
These targets were faint ($b_J > 19$ on the UKST IIIaJ plates) 
and required long integration times of several hours per object at this
high dispersion.

\begin{figure}[t]
%\vspace{6cm}
\centerline{\psfig{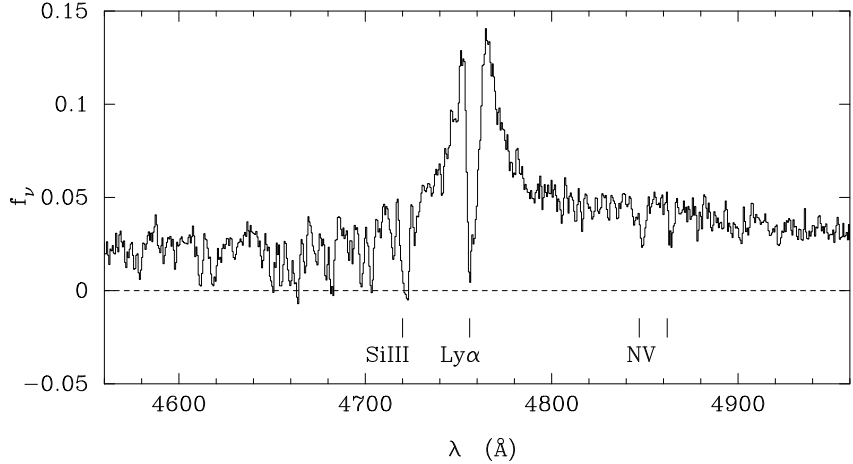} }
\caption[spectra]
{Spectrum of the Ly-$\alpha$ region  at 1-\AA\ resolution for the $z=2.914$
CSS quasar MQS\,0246$-$231. Absorption lines are marked. 
}
\label{q0246}
\end{figure}
 
In all three quasars, strong Ly-$\alpha$ absorption is seen within 2000\,\kms\ of
the quasar redshift.   The Ly-$\alpha$ absorption lines are strong but 
not damped, making estimates of the column density uncertain. 
Typical column densities of $N_{\rm HI} \sim 10^{14} - 
10^{18}$\,cm$^{-2}$ are indicated along the line of sight. 
Dust is undoubtedly present as well, which may have increased the
Ly-$\alpha$ optical depth via scattering.  The Ly-$\alpha$ 
absorption lines are both blue- and red-shifted with respect 
to the Ly-$\alpha$ emission-line peak, and appear broadened 
with FWHMs of 100--400\,\kms. Blue-shifted absorption systems are most
common, perhaps indicating outflowing clouds from the nucleus. 
The presence of at least one redshifted absorption system 
relative to the quasar emission redshift, and the broadened 
lines, favours an intrinsic origin for the absorbing material 
rather than absorption by intervening galaxies.

Over the wavelength range observed, 
 many other species are visible, ranging from N\,{\sc v} to 
O\,{\sc i}.  The variety of species present indicates that the absorbing
clouds span a range of density and ionisation. One quasar shows 
high-ionisation lines (N\,{\sc v} and Si\,{\sc iii}) along with 
\lya\ in the same system,  another quasar both high- and low-ionisation 
species (Si\,{\sc ii}, O\,{\sc i}, C\,{\sc ii}). In the third spectrum, 
the Ly-$\alpha$ system lying closest to the quasar redshift contains no 
obvious metal 
lines,  but a second absorption system containing \lya\ as well as strong 
low ionisation lines is present 5000\,\kms\ to the blue of the quasar
redshift. 

Figure \ref{q0246} shows an example of a spectrum of a CSS quasar at
$z=2.914$, MQS\,0246$-$231. This quasar is very faint --- $b_{\rm J} 
= 21.5$ on the UKST IIIaJ plates. Strong Ly-$\alpha$ absorption is clearly 
seen along with the high-ionisation absorption lines of N\,{\sc v} and 
Si\,{\sc iii}, which are marked. 
There is a slight redshift mis-match between \lya, N\,{\sc v} and
Si\,{\sc iii}, which probably indicates that the
Ly-$\alpha$ absorption arises in spatially distinct regions. All four
absorption lines are clearly resolved, indicating a relatively large velocity
dispersion for the clouds (up to 400\,\kms).  The Ly-$\alpha$
emission-line profile in this quasar is relatively narrow 
(1000\,\kms\ FWHM), a width more typical of radio galaxies.

\section{Parallels between CSS quasars and HZRGs}

The strong Ly-$\alpha$ absorption found towards CSS quasars is very similar 
to that reported towards a number of high-redshift radio galaxies (HZRGs)
(e.g. van Ojik et al. 1997). In HZRGs, the absorbers are probably 
related to extended kpc-scale \lya\ emission 
regions, which indicate that reservoirs of H\,{\sc i} gas envelope 
the radio source. The Ly-$\alpha$ absorbers in HZRGs are observed to cover 
the entire area of extended Ly-$\alpha$ emission, reaching tens of kpc from the
nucleus. The absorbers in HZRGs are also metal-enriched --- species 
including C\,{\sc iv} and Si\,{\sc iii} accompany the Ly-$\alpha$
absorption. In HZRGs, Ly-$\alpha$ absorption occurs 
preferentially towards sources of small linear size (van Ojik et al. 1997) 
and with bent radio structures (Barthel \& Miley 1988), suggesting a 
link between absorption and radio source size and also direct interactions 
between the H\,{\sc i} halo and the radio jets.

\begin{figure}[t]
\centerline{\psfig{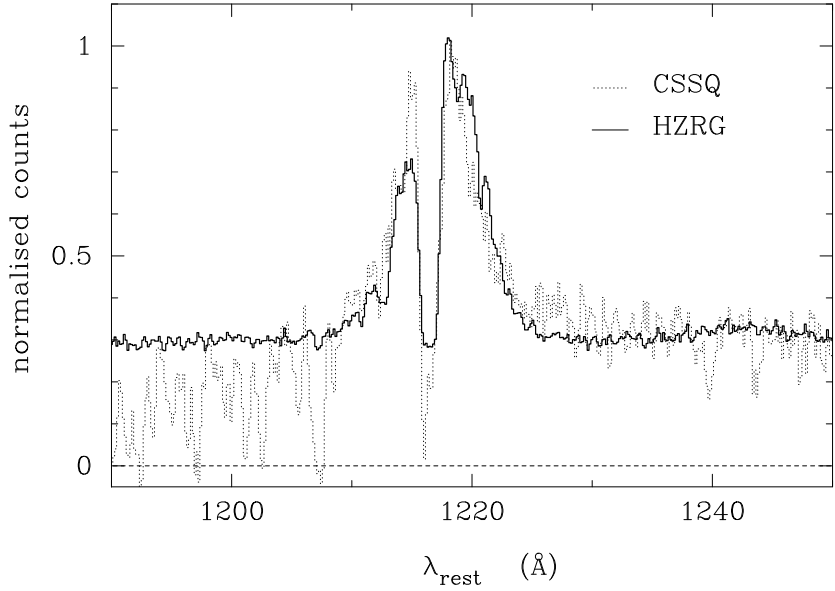} }
\caption[spectra] 
{ \small Overlaid restframe spectra (1\AA\ resolution) 
of a MQS CSS quasar (dotted line) and HZRG (MRC0943$-$242 
from R\"ottgering et al. 1995) (solid line)
at $z=2.9$.  The HZRG spectrum has been shifted upwards to 
match the quasar continuum level and then scaled to the \lya\ 
emission peak. Note the remarkable correspondence between the 
emission and absorption profiles. 
}
\label{two}
\end{figure}

In fact many of the above characteristics of HZRGs are shared with CSS sources, 
raising the possibility that HZRGs and CSS sources are intrinsically 
similar objects.

In Figure \ref{two}, the 1-\AA\ resolution Ly-$\alpha$ spectrum of a 
HZRG at $z=2.9$, MRC\,0943$-$242 (R\"ottgering et al. 1995), 
has been scaled and plotted on top of \lya\ for the CSS quasar 
MQS\,0246$-$231 (from Figure \ref{q0246}).  The similarities in 
the Ly-$\alpha$ absorption and emission properties are striking. The only
obvious difference between the two is the presence of continuum emission 
in the quasar. 

This example strongly supports the idea that many HZRGs
are in fact misaligned CSS quasars with their continuum and broad-line 
regions hidden by a dusty torus. Such a picture is entirely 
consistent with the inference of hidden quasars in HZRGs 
through spectropolarimetry (e.g. Cimatti et al. 1997) 
and the presence of continuum and emission-line light 
aligned with the radio axis in many HZRGs (e.g. Rush et al. 1997).

\section{Discussion}

By noting the similarities between HZRGs and CSS quasars, the properties
of intrinsically small radio sources can begin to be unified.
This spectroscopic  study shows that some of the properties 
of HZRGs are in fact common characteristics of small radio sources, including 
many CSS sources. Whether these properties are a consequence of youth
remains to be confirmed --- some fraction of older, frustrated sources is
still expected. The similarities between HZRGs and CSS quasars places
even more emphasis on the study of CSS sources over a wide range of 
redshifts.

The presence of Ly-$\alpha$ absorption with \zemzabs\ preferentially in sources
of small linear size suggests that the absorption properties of quasars
are linked directly to the evolution of the radio source. It can 
be postulated that radio sources are born in cocoons of enriched H\,{\sc i} 
gas which is ionised and pushed aside as the radio source grows. This
explains the lower rate of \zemzabs\ absorption in larger sources. The
origin of the H\,{\sc i} cocoons, however, and their possible 
r\^ole in triggering radio emission in active galaxies, remains 
unknown. Remnants of past galaxy mergers and reservoirs of 
primordial gas have both been postulated as explanations for the
Ly-$\alpha$ haloes around HZRGs.

\section{Conclusions}

Associated (\zemzabs) Ly-$\alpha$ absorption is common in CSS quasars. 
The absorbers may be metal enriched, and span a range of density 
and ionisation. The properties of these Ly-$\alpha$ absorbers appear 
identical to those seen towards HZRGs, especially ones of small 
linear size.  On the basis of their similar H\,{\sc i} environments, as
probed by absorption, and other evidence,  it is likely that 
many HZRGs and CSS quasars are intrinsically similar objects. 

The properties of small radio sources are crucial for investigating the
processes which drive radio source evolution. Associated absorption in
particular appears to offer a new way of probing possible 
intrinsic differences between large and small radio sources.

\end{document}